\documentclass[lettersize,journal]{IEEEtran}
\usepackage{amsmath,amsfonts}
\usepackage{array}
\usepackage[caption=false,font=normalsize,labelfont=sf,textfont=sf]{subfig}
\usepackage{textcomp}
\usepackage{stfloats}
\usepackage{url}
\usepackage{verbatim}
\usepackage{graphicx}
\usepackage{graphics}
\usepackage{algpseudocode}
\usepackage{color}
\usepackage{comment}
\usepackage{setspace} 
\usepackage{hyperref}
\usepackage{footnote}
\usepackage{lipsum}
\usepackage{multirow}
\def\BibTeX{{\rm B\kern-.05em{\sc i\kern-.025em b}\kern-.08em
    T\kern-.1667em\lower.7ex\hbox{E}\kern-.125emX}}
\usepackage{balance}
\begin{document}
\title{Protected Working Groups-based Resilient Resource Provisioning in MCF-enabled SDM-EONs}
\author{Anjali Sharma, Varsha Lohani, and Yatindra Nath Singh\\
\textit{Electrical Engineering Department, Indian Institute of Technology Kanpur, India}}

\maketitle

\begin{abstract}
Space Division Multiplexed- Elastic Optical Networks using Multicore Fibers are a promising and viable solution to meet the increasing heterogeneous bandwidth demands. The extra capacity gained due to spatial parameters in SDM-EONs could encounter detrimental losses if any link fails and timely restoration is not done. This paper proposes a Protected and Unprotected Working Core Groups assignment (PWCG/ UPWCG) scheme for differentiated connection requests in multicore fiber-enabled SDM-EONs. A PWCG is inherently protected by resources in a Dedicated Spare Core Group (DSCG). First, we divide the cores into three groups using traffic and crosstalk considerations. In the second step, we use the obtained core groups for resource provisioning in dynamic network scenarios. The effectiveness of our proposed technique is compared with a Link Disjoint Path Protection (LDPP) technique, and the simulation study verifies our assertions and the findings. 
\end{abstract}


\section{Introduction}
Recently, Multicore Fibers (MCF) have been the centre of research as they have the same attenuation profile as a single-mode fiber and each core can carry signals independently giving advantage of parallelism \cite{C1_SDM_Survey1}. The need to transmit more information in a single fiber led to significant research in multicore fibers. The transmission capacity can be increased linearly within the existing fiber dimension by increasing the number of spatial parameters or cores resulting in MCF. However, the number of cores or spatial parameters in a single fiber is limited by the core pitch (distance between centre of two adjacent cores). The smaller core pitch value leads to more crosstalk impairment, where the power of the signal in one core is leaked into the adjacent cores and deteriorates the performance of the adjacent core's transmission. There is an inherent trade-off between the number of usable spectrum slices in the adjacent cores and the loss of signal quality due to crosstalk accumulation. Crosstalk-aware and crosstalk-avoidance techniques are used when providing resources to connection requests to mitigate the severe effects of crosstalk in Multicore Fibers-based SDM-EONs (MCF-SDM-EON).\\
When performing the Crosstalk-Avoidance or Crosstalk-Aware Resource provisioning in Route, Modulation, Core and Spectrum Assignment (RMCSA), a fraction of cores' equivalent bandwidth becomes unusable due to intercore-crosstalk considerations. The maximum spectral resource utilization is limited to 60-70$\%$ because of these considerations and constraints in RMCSA. The excess capacity gain of the SDM-EONs can be truly realized by utilizing the crosstalk-affected capacity also for some purpose. One of the possible techniques is to design such fibers, which only have the usable capacity and are unaffected by the crosstalk impairments. Another technique is to use the available capacity judiciously and effectively. One such efficient and reliable technique is to use the affected capacity to provide resilience to the operating set of connection requests. In that way, the previously unused capacity can be useful, even if for a short duration during the failure time but at the cost of compromised signal quality. The previously proposed protection techniques have also used similar techniques, but we have tried to formulate a much simpler protection technique based on the core utilization and traffic characteristics.\\  
Resilience techniques are employed with provisioning to avoid data losses in the event of failure of the network components. The amount of data that a single MCF fiber can carry is much more than a single-core fiber, and a proportional loss of bandwidth is expected due to the failure. The protection techniques need to be employed to provide resilience, where the spare capacities are reserved for the backup. The active connections are switched over to spare capacity resources in case of a failure \cite{C7_I1}. The protection technique could be dedicated-type or shared-type. In shared-type protection, multiple primary routes share the same spare capacity for backup and primary routes need to be link-disjoint. The backup routes are calculated using the spare capacity, and 100$\%$ re-establishment of affected connections is possible. The amount of backup resources needed depends on how many failures atmost can happen at a time. The objective of the resilience techniques is to provide 100$\%$ protection with minimum spare capacity and restoration time. The reserved spare capacity creates redundancy and may reduce the effective usable capacity. \\
Several works have considered crosstalk-aware resource provisioning together with the protection techniques. Authors in \cite{C7_P1}  have used Failure-Independent Path Protecting pre-configured cycles to provide resilience using a PERFECTA algorithm. Using pre-configured cycles give the advantage of shorter restoration time and higher spare capacity efficiency. Authors in \cite{C7_P2} and \cite{C7_P3} have used Shared Backup Path Protection (SBPP) to provide 100$\%$ protection together with minimal crosstalk effects and blocking of the connections. In \cite{C7_P4}, authors have proposed a Link-disjoint Path Protection (LDPP) technique with crosstalk, blocking and spectrum utilization considerations. Authors in \cite{C7_P5} and \cite{C7_P6} have used the multipath strategy in combating the losses due to single link failure. Many proposed works have yet to discuss spare capacity redundancy and its associated leverage with crosstalk awareness. In this work, we propose to use the crosstalk-affected core capacities as the reserved spare capacity for protecting connection requests.

\section{Proposed Work}
\begin{figure}[ht]
\begin{center}
\includegraphics[width=\linewidth,keepaspectratio]{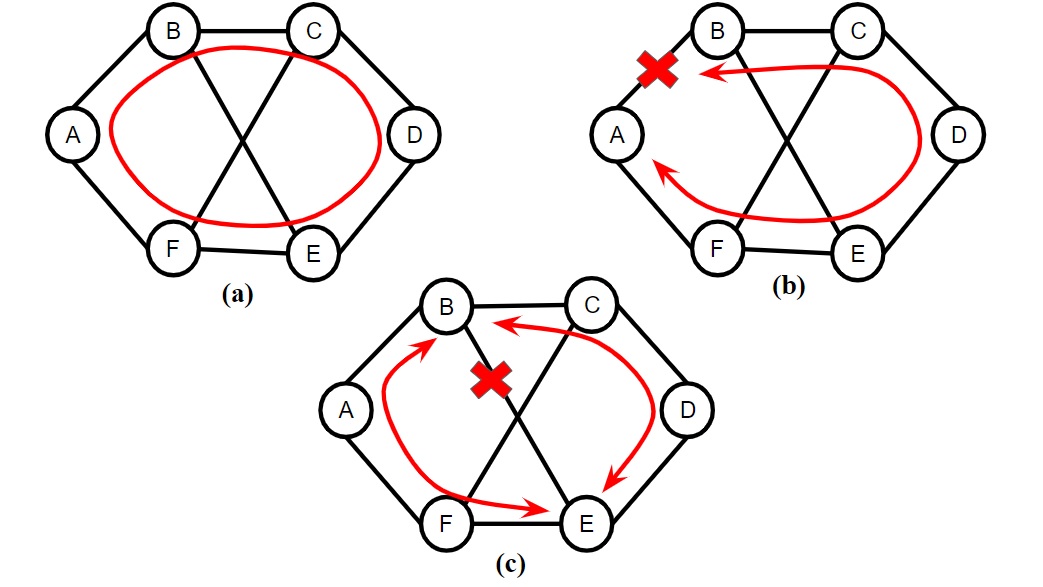}
\caption{(a) Protection Cycle A-B-C-D-E-F-A, (b) when  A-B fails (undirected), the services are protected by providing capacity on a single backup route, B-C-D-E-F-A, created by surviving links of the protection cycle, and (c) when B-E fails (undirected), the services are protected by providing capacity on two backup routes, B-A-F-E and B-C-D-E, created by surviving links of the protection cycle.   }
\label{fig:C7_protection_cycles}
\end{center}
\end{figure}

In this work, we leverage the unusable capacity due to the crosstalk considerations as the spare capacity needed to protect the prioritized services in SDM-EONs. The concept of the Protected Working Cores (PWC) group has been derived from the Protected Working Capacity Envelope-based protection in WDM to provide link protection. In place of wavelength-level protection switching, we are using Core-level protection as it results in simplified and cost-efficient switching operations and timely restoration, \cite{C7_I2}. This resilience model contains three types of resources in the network links: Unprotected, Protected Envelope and Spare Resources. We adopt a similar approach, and the MCF's cores are divided into three groups: Dedicated Spare Cores (DSCG), Protected Working Cores (PWCG), and Unprotected Working Cores (UPWCG) groups. The connection requests in the PWCGs are inherently protected by the protection structures using the DSCGs. When any failure occurs, the connections in the PWCGs are switched to the backup routes in DSCGs. The backup routes are in the form of protection cycles, which can provide better spare capacity efficiency. The protection cycles can protect an on-cycle failed link by providing one backup route and a failed straddling link by providing two backup routes, as shown in Figure ~\ref{fig:C7_protection_cycles}. The full waveband and full core switching are assumed here. The connection requests provisioned in the UPWCG require re-routing of the connections using the leftover capacities after PWCG connections are protected in the event of failure. There is no guarantee of restoration, and the time to find backup resources may also keep increasing, leading to the deterioration of the quality of service. In addition to protection cycles, other protection structures could also be used. \\
We have used three Service Class Types (SCT) to assign the resources in the mentioned core groups. Table 6.1 shows the type of protection required by each SCT and the core group to which they belong. Considering the cost, the SCT-I are the service type requiring the best and uninterrupted service quality and could be priced more than the other SCTs. 

\begin{table}[ht]
\caption{Service Class Types}
\begin{center}
\begin{tabular}{ |c|c|c| } 
 \hline
 \textbf{Service Class}  & \textbf{Quality of} & \textbf{Provisioning}\\
 \textbf{Type (SCT)} & \textbf{Protection} & \textbf{Core Group}  \\ 
 \hline
   SCT-I & 100$\%$ Protection & PWCG \\ 
 \hline
  SCT-II & Best Effort & UPWCG  \\
   & Restoration &  \\
 \hline
  SCT-III & No Protection & UPWCG and DSCG based on  \\
  & & effective protection\\
 \hline
 \end{tabular}
\end{center}
\label{tab:C6_tab1}
\end{table}
When SCTs' connection requests arrive and depart dynamically in the network, we must find the most relevant resources and accommodate the maximum number of such requests. The whole process of providing resilient resource provisioning requires grouping the cores first. Then the connection requests are assigned to their relevant core groups. We divide the process into three parts: Resource Planning, Resource Provisioning and Reprovisioning after a failure event. \\

\begin{table}[ht]

\caption{List of Parameters and Variables in the ILP model.}
\begin{center}
\resizebox{0.95\linewidth}{!}{
\begin{tabular}{|c|c|}
\hline

\multicolumn{2}{|c|}{Sets}  \\
 \hline
V & Set of Nodes\\
L & Set of Links\\
P & Set of Protection Cycles\\
K & Set of Cores on each link\\
\hline
\multicolumn{2}{|c|}{Parameters}  \\
\hline
CP & Cost of Protected Working Groups\\
CU & Cost of Unprotected Working Groups\\
CS & Cost of Dedicated Spare Working Groups\\
$XTC_k$ & Cost of $k^{th}$ core due to Crosstalk-considerations\\
$W_{max}$ & Maximum protected core groups required \\
$W_{min}$ & Minimum protected core groups required \\
$\pi_{i}^{p}$ & 1, if $p^{th}$ protection cycle crosses link $i$, otherwise 0\\
$X_{i}^{p}$ & 1, if $p^{th}$ protection cycle provides on-cycle protection to $i^{th}$ link, 2 for straddling \\
 & protection, 0 otherwise\\
 \hline
\multicolumn{2}{|c|}{Variables}  \\
\hline
$S_i$ & Total spare capacity on $i^{th}$ link\\
$P_i$ & Total protected capacity on $i^{th}$ link\\
$UP_i$ & Total unprotected capacity on $i^{th}$ link\\
$n_p$ & Number of copies of $p^{th}$ protection cycle\\
$SL_{i}^{k}$ &1 if $k^{th}$ core on $i^{th}$ link is assigned DSCG, 0 otherwise \\
$PL_{i}^{k}$ &1 if $k^{th}$ core on $i^{th}$ link is assigned PWCG, 0 otherwise \\
$UPL_{i}^{k}$ &1 if $k^{th}$ core on $i^{th}$ link is assigned UPWCG, 0 otherwise \\
$\delta_i^k$ & the cost for $k^{th}$ core in $i^{th}$ link \\
\hline
\end{tabular}
}
\label{tab:C6_tab2}
\end{center}
\end{table}
\subsection{Resource Planning Model}
Resource Planning requires prior information on types of arriving connection requests, i.e., the fraction of total arriving connection requests constituting SCT-I, SCT-II and SCT-III individually. The total capacity available in each MCF-link is equivalent to the number of cores as we have considered core-level granularity for the protection structures. We have used an Integer Linear Programming (ILP) formulation to find the most optimum core group arrangement to provide resilient resource provisioning in the next step. \\
We represent the SDM-EON as a mathematical model using the ILP formulation and notations. The network is modelled as an graph with \textit{V} number of nodes and \textit{L} number of bi-directional links. In this model, capacity of the links is represented by the number of cores \textit{K}. To provide the backup routes/paths, we have used the \textit{P} number of protection cycles. Table 6.2 lists the sets, parameters and variables needed for the ILP model. The ILP is formulated as follows.\\

\begin{equation}
             \textnormal{\textbf{Minimize}}   \sum_{i \in L, k \in K} XTC_k * \delta_i^k            
\end{equation}
\textbf{Subject to:}
\begin{equation}
             S_i = \sum_{p \in P} \pi_i^p * n_p , \forall i \in L           
\end{equation}
\begin{equation}
             P_i = \sum_{p \in P} X_i^p * n_p , \forall i \in L 
\end{equation}
\begin{equation}
             UP_i = K-P_i -S_i , \forall i \in L 
\end{equation}
\begin{equation}
             W_{max} = max(P_i) , \forall i \in L 
\end{equation}
\begin{equation}
             W_{min} = min(P_i) , \forall i \in L 
\end{equation}
\begin{equation}
             \sum_{k \in K} SL_i^k = S_i, \forall i \in L 
\end{equation}
\begin{equation}
             \sum_{k \in K} PL_i^k = P_i, \forall i \in L 
\end{equation}
\begin{equation}
             \sum_{k \in K} UPL_i^k = UP_i, \forall i \in L 
\end{equation}
\begin{equation}
             PL_i^k + UPL_i^k + SL_i^k = 1, \forall i \in L, \forall k \in K 
\end{equation}
\begin{equation}
             \delta_i^k = (CS*SL_i^k) + ( CU*UPL_i^k) + (CP*PL_i^k) , \forall i \in L, 
             \forall k \in K
\end{equation}
A connection request requiring 100$\%$ protection should be ready to pay extra in terms of cost-per-bit for reliable service. The under-utilized cores are costly to maintain from the service provider's point of view. So, the overall objective is to divide the cores among the three categories to minimize the overall cost, eq. 1. The constraint in eq. 2 assigns the spare capacities (cores) and eq. 3 gives the protected capacities on each link of the network. Eq. 4 gives the total capacity combination on each link. Eq. 5 and Eq. 6 gives the variation of the protected working capacity on each link. Eq. 7, Eq. 8 and Eq. 9 gives the assignment of each core on the links for spare, protected and unprotected capacities, respectively. Constraint in Eq. 10 ensures that each core is assigned to exactly one core group during the group assignments. Finally, eq. 11 gives the core cost status after grouping them on each link.   
\subsection{Resource Provisioning}
\begin{figure}[ht]
\begin{center}
\includegraphics[width=\linewidth,keepaspectratio]{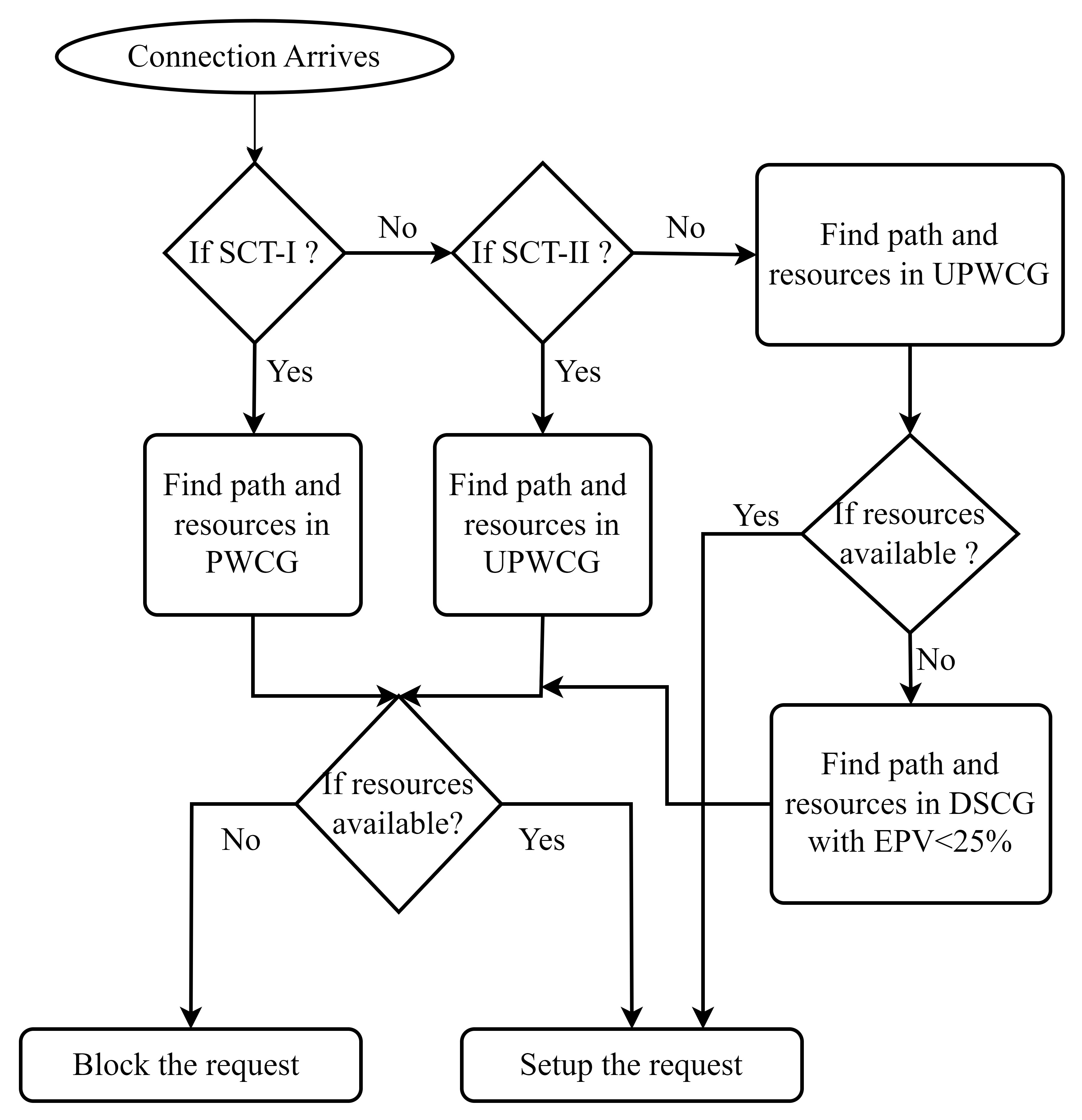}
\caption{Flowchart of the resource provisioning with PWC groups}
\label{fig:C7_fig3}
\end{center}
\end{figure}
In the resource provisioning step, the connection requests arrive and depart dynamically. The shortest route is selected using the distance as the cost. On the selected route, the resources are chosen from the relevant core groups using first-fit spectrum slices and crosstalk considerations; if not available, the connection requests are blocked. If the arrived connection request is of type SCT-I, then the resources are searched in PWCG. If the arrived connection request is of type SCT-II, then the resources are searched in UPWCG. If the resources are available in the PWCG and UPWCG for SCT-I and SCT-II, respectively, then the connection is provisioned and if not, then blocked. For connection requests of type SCT-III, when the connection request arrives, the resources are searched for in UPWCG. If available, a connection request is provisioned. If not, the Effective Protection Values (EPV) of the cores in the DSCG are calculated. EPV is the ratio of the spectrum slices used to provide resilience and the total number of slices in the core. If the EPV is less than 25$\%$ of A-priori protection value (spare capacity reserved for protected capacities), we provision the SCT-III in the DSCG. If not, the connection request is blocked. The flowchart of the resource provisioning is shown in Figure ~\ref{fig:C7_fig3}.  \\

If a single link failure occurs, the active connections on the failed links are reconfigured according to their protection type. At first, all the SCT-III occupying DSCG gets released and the resources are released. The SCT-III are not protected; hence, all the affected requests of this type are blocked. The connection requests of SCT-I are switched to the same spectrum slices in the DSC protection structure, which will be configured at that time to make the backup path. The connection requests of type SCT-II are provided backup resources on the fly, i.e., the backup resources are calculated using the UPWCGs of the surviving links to provide link protection. \\  

\begin{figure*}[ht]
\begin{center}
\includegraphics[width=0.7\linewidth,keepaspectratio]{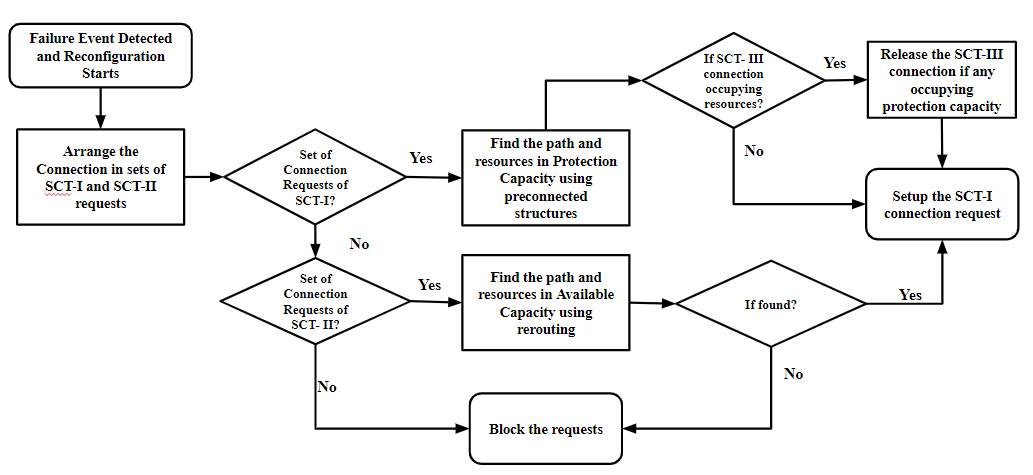}
\caption{Schematic of the resource provisioning with PWC groups}
\label{fig:C7_fig4}
\end{center}
\end{figure*}
The computational complexity of the proposed technique involves only the routing and spectrum assignment steps, $O(V^2 + K + V.S)$, where V is the number of nodes in the network, K is the number of cores, and S is the number of spectrum slices on each core of the link. The ILP is performed only once in the offline scenario, which is why the computation time of ILP is neglected.\\

\section{Simulation Setup}
\begin{figure}[ht]
\begin{center}
\includegraphics[width=0.6\linewidth,keepaspectratio]{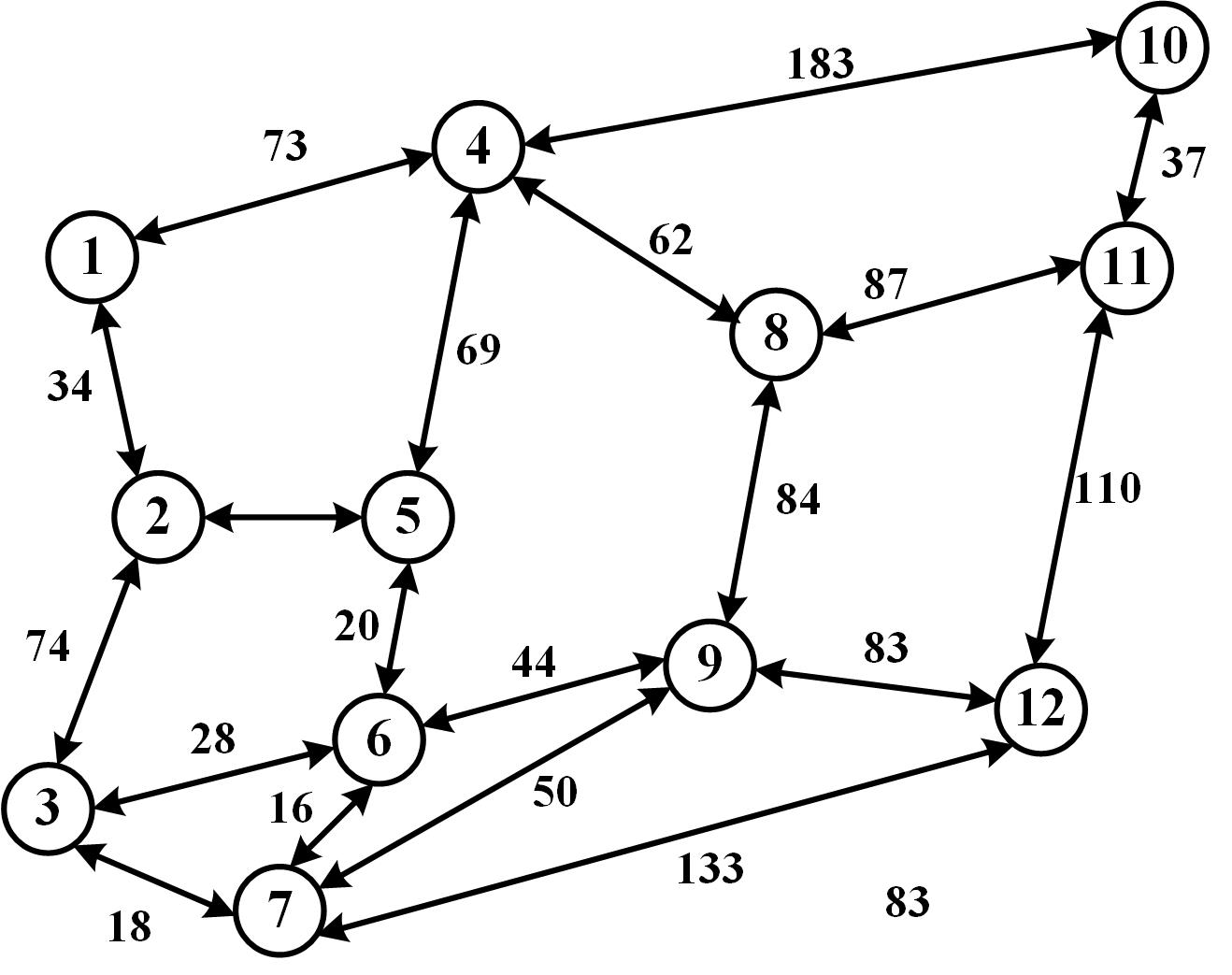}
\caption{Test Network}
\label{fig:C7_net}
\end{center}
\end{figure}
We simulated the proposed and benchmark techniques on a Finland-12 network. The network comprises of 12 nodes and 19 MCF links, as shown in Figure ~\ref{fig:C7_net}. Each MCF link has seven cores, and each core has 320 spectral slots. Here also, the spectral slice granularity is 12.5 GHz. The cost parameters used in ILP are shown in Table ~\ref{tab:C6_tab3}. The connection arrival is Poisson-distributed, and the connection service time is exponentially distributed. The bandwidth demand ranges from 50 Gbps to 250 Gbps, and the required spectrum slices are calculated accordingly. Appropriate modulation levels are considered, ranging from BPSK up to 64-QAM. The probability of the arrival of SCT-I, SCT-II and SCT-III are equal. The simulations are run over 20 iterations for 100000 connection requests in a steady-state scenario. A single link failure event is considered, and the time of failure is estimated using the individual link's Mean Time to Failure (MTTF) values. \\
We have compared the proposed resilient resource provisioning technique with a Link-Disjoint Path Protection (LDPP). In the LDPP, a connection request is provisioned with two routes, primary and backup routes, using the shortest paths and spectral resources available on it.\\
\begin{table}[ht]

\caption{ILP model Parameter Values}
\begin{center}
\begin{tabular}{|c|c|c|c|c|}
\hline

CP & CU & CS & $W_{max}$ & $W_{min}$ \\
\hline
3 & 2 & 1 & 2 & 1 \\
 \hline
\end{tabular}
\label{tab:C6_tab3}
\end{center}
\end{table}
Connection Blocking Ratio (CBR), Bandwidth Blocking Ratio (BBR), Affected Connection Requests (ACR), Affected Bandwidth (ABW), Protection Level and Redundancy are the performance indicators for the proposed technique (PWCG) and the benchmark technique. BBR is the ratio of the amount of blocked bandwidth to total arrived bandwidth requests in the network during simulation. ACR is the average number of connection requests being affected after the occurrence of failure and requiring reprovisioning. ABW gives the total affected bandwidth (in spectrum slices) in active connection requests due to a single link failure. The Protection Level is the percentage of active connection requests protected after the single link failure and given the reconfigured backup routes. Redundancy is the ratio of the reserved spare capacity to the working capacities during the simulation.  \\
\section{Performance Evaluation}

\begin{figure}[ht]
\begin{center}
\includegraphics[width=\linewidth,keepaspectratio]{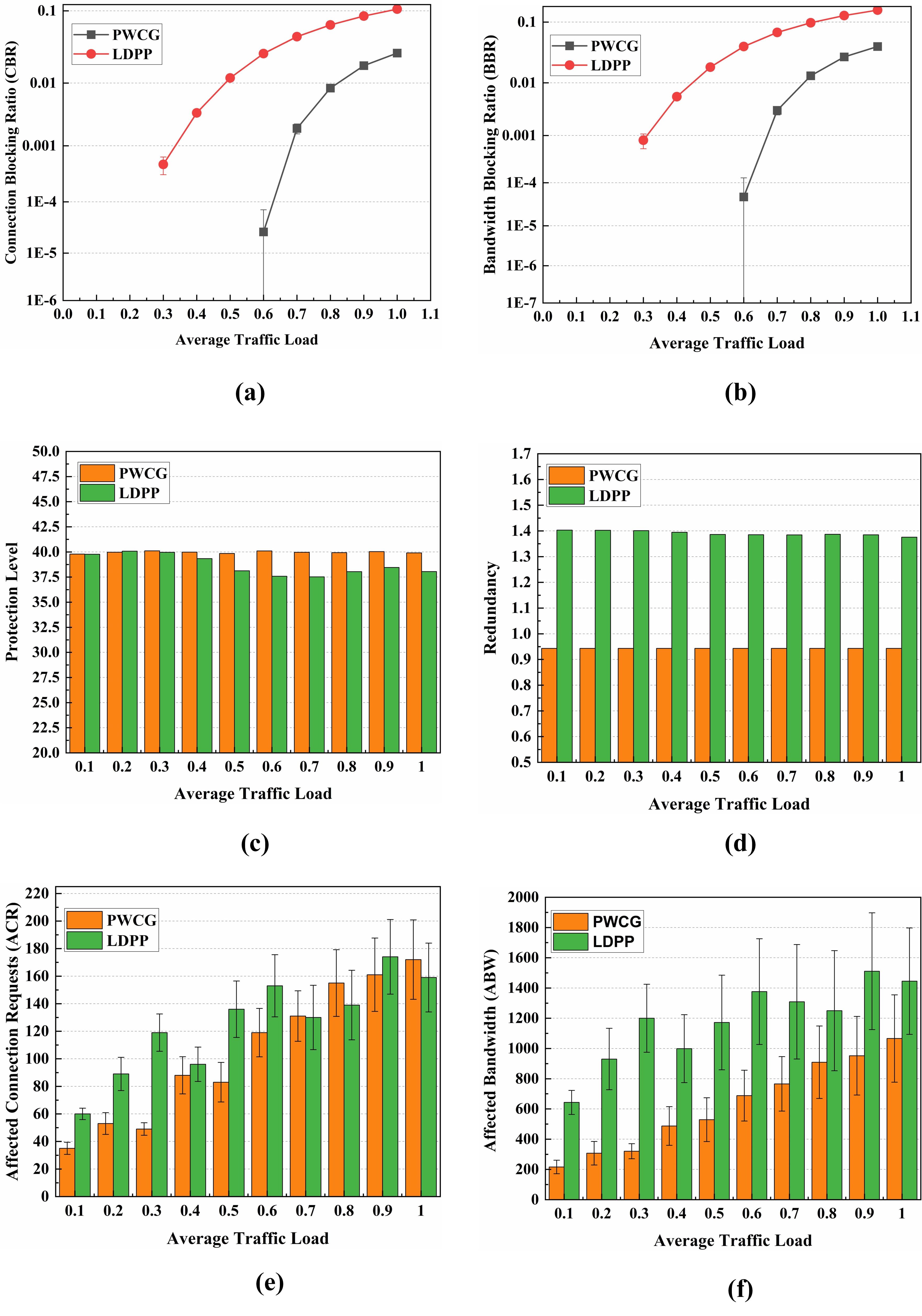}
\caption{Performance Parameters vs Normalized Traffic Load}
\label{fig:C7_perf}
\end{center}
\end{figure}
Figure ~\ref{fig:C7_perf} shows the different measures for network performance. Figure ~\ref{fig:C7_perf}(a) and Figure ~\ref{fig:C7_perf}(b) record the traffic admissibility of the two techniques using CBR and BBR. The proposed group-based protection, PWCG, performs better than the LDPP technique. The LDPP technique occupies resources on two routes, the shortest route and the second shortest route, with spectrum slices calculated using appropriate modulation levels. Thus, more blocking is witnessed in the LDPP technique due to the unavailability of spectral resources due to 50$\%$ more redundancy. The blocked services (requests and bandwidth) are within 5$\%$ of the total arrived connections in the proposed technique. Figure ~\ref{fig:C7_perf}(c) and Figure ~\ref{fig:C7_perf}(d) record the protection level and the redundancy required by two protection techniques. LDPP has almost 50$\%$ more redundancy than PWCG, which means backup routes use more resources than the primary working route in LDPP. At higher traffic load values, many connection requests are provisioned without protection due to higher redundancy caused by protected services in LDPP (Figure ~\ref{fig:C7_perf}(c)). Due to proper resource grouping, the protection level is more in the PWCG technique. \\
Figure ~\ref{fig:C7_perf}(e) and Figure ~\ref{fig:C7_perf}(f) record the impact of single link failures. After the single link failure, the protected connection requests are restored by signalling and switching to backup routes. The unprotected connections of SCT-II, find the backup resources on the fly using complex signalling steps and surviving spectral resources. But, the unprotected connections of SCT-III lose the resources throughout the working route. ACR and ABW values of the LDPP technique are more than the PWCG technique, as most of the active connections are left unprotected due to redundancy.\\

\section{Conclusion}
In this proposed work, we have considered under-utilized resources due to crosstalk considerations in the MCF of SDM-EONs to provide resilient resource provisioning. The centre-most cores in the circular core arrangement of MCF are the worst affected. We divide the cores into three groups depending on their survivability characteristics using ILP model: Protected, Unprotected and Spare Core Groups. These groups can provide provisioning to three classes of services.\\
We compared the performance of our Protection Cycles-based PWCG with a Link-disjoint dedicated path protection. The simulation results support our assertions that the proposed technique provides better protection levels and survivability in the event of a single link failure. It also achieves lower blocking of the arriving connection requests than the benchmark technique, making it a resilient provisioning technique.

   \end{document}